\documentstyle[preprint,floats,tighten,prd,aps]{revtex}
\input epsf.tex
\def\DESepsf(#1 width #2){\epsfxsize=#2 \epsfbox{#1}}
\begin{document}
\draft

\title{Choosing integration points for QCD calculations by
numerical integration}
\author{ Davison E.\ Soper}
\address{Institute of Theoretical Science, 
University of Oregon, Eugene, OR 97403}
\date{23 March 2001}
\maketitle

\begin{abstract}
I discuss how to sample the space of parton momenta in order to best
perform the numerical integrations that lead to a calculation of three
jet cross sections and similar observables in electron-positron
annihilation.
\end{abstract}

\pacs{}


\section{Introduction}
\label{sec:introduction}

There is an important class of computer programs that do calculations in
quantum chromodynamics (QCD) in which the calculation is performed at
next-to-leading order in perturbation theory and allows for the
determination of a variety of characteristics of the final state.
This paper concerns a significant technical issue that arises in such
programs when a ``completely numerical'' integration algorithm is used:
how should one choose the integration points?

I consider the calculation of ``three-jet-like'' observables in $e^+
e^-$ annihilation. A given program in the class considered can be used
to calculate a jet cross section (with any infrared safe choice of jet
definition)  or observables like the thrust distribution. Such a program
generates random partonic events consisting of three or four final state
quarks, antiquarks, and gluons. Each event comes with a calculated 
weight. A separate routine then calculates the contribution to the
desired observable for each event, averaging over the events with their
weights.

The weights are treated as probabilities. However, these weights can be
both positive or negative. This is an almost inevitable consequence of
quantum mechanics. The calculated observable is proportional to the
square of a quantum amplitude and is thus positive. However, as soon
as one divides the amplitude into pieces for purposes of calculation,
one finds that while the square of each piece is positive, the
interference terms between different pieces can have either sign. Thus
the kind of program discussed here stands in contrast to the tree-level
event generators in which, by simplifying the physics, one can
generally arrange to have all the weights be positive, or, even, to
be all equal to 1.

To understand the algorithms used in the class of programs described
above, it is best to think of the programs as performing
integrations over momenta in which the quantum matrix elements and the
measurement functions form the integrand. There are two basic algorithms
for performing the integrations. The older is due to Ellis, Ross, and
Terrano (ERT) \cite{ERT}. (The first general purpose implementation of
this method for three-jet-like observables in $e^+ e^-$ annihilation
was that of Kunszt and Nason \cite{KN}). In this method, some of the
integrations are performed analytically ahead of time. The other
integrations are performed numerically by the Monte Carlo method. The
integrations are divergent and are regulated by analytical
continuation to $3 - 2\epsilon$ space dimensions and a scheme of
subtractions or cutoffs. The second method is much newer
\cite{beowulfPRL,beowulfPRD}. In this method, all of the integrations
are done by Monte Carlo numerical integration. With this method, the
integrals are all convergent (after removal of the ultraviolet
divergences by a straightforward renormalization procedure). 

Since this method is quite new, one cannot yet say for what problems it
might do better than the now standard ERT method. I can point out that
the current incarnation of the numerical method has convergence
problems for three jet quantities that receive important contributions
from final states that are close to the two jet limit. For example, one
gets good results for the three jet cross section or for the thrust
distribution away from $T = 1$, but poorly converging results for the
energy-energy correlation function. On the other hand, the numerical
method offers evident advantages in flexibility to modify the
integrand. Thus one should be able to attack problems that are not
accessible to the ERT method.

The numerical integration method exists as computer code
\cite{beowulfcode} with accompanying technical notes
\cite{beowulfnotes} and many of the basic ideas behind it have been
described in the two papers \cite{beowulfPRL,beowulfPRD}. The present
paper is devoted to an exposition of the choice of integration points
needed for this method. This subject may seem unimportant.
Furthermore, it may seem to be a just a part of the numerical
algorithms branch of computer science and thus to be uninteresting
from the point of view of physics. However, practitioners of the art
of numerical integrations in physics know that the choice of
integration points is extremely important. If everything else is
perfect but the integration points are badly chosen, the computer
will, like Sisyphus, be faced with an unending task because the
precision of the answer will not improve as the computational time
increases. Furthermore, the problem is a problem in physics. All that
the theory of numerical integration tells us is that the density of
integration points should be well matched to the structure of the
function to be integrated. The function to be integrated comes from
Feynman diagrams, so we are really faced with the problem of
understanding the structure of the quantum scattering processes.
Perhaps surprisingly, the aspects of quantum scattering that are
important in numerical method are completely different than the
aspects that are important in the ERT method.

In Sec.~II below, I briefly review the basics of the numerical
integration method with the aim of setting the notation and making it
possible to read this paper independently of the previous two papers.
Then, in the sections that follow, I turn to the main issue of this
paper, the choice of integration points. I try to keep this discussion
succinct. The main features of the method are covered, but
implementation details are left to the program code and its accompanying
technical notes \cite{beowulfcode,beowulfnotes}.

The exposition begins in Sec.~III with a review of the general
principles of Monte Carlo integration as they apply to this
calculation. Sec.~IV is the heart of the paper and covers the
organization of the method used to choose the most integration points
in the most important integration regions. Part of the method used
considers the possible three parton final states. This part is
explained in Sec.~V.  The other parts of the method concern the
singularities of virtual graphs and their relation to the energy
denominators in 2 parton to $n$ parton scattering.  The various
possibilities are covered in Secs.~VI through IX.  Of special
importance is the elliptical coordinate system introduced in Sec.~VI. 
A brief summary is presented in Sec.~X.

\section{Review of the numerical method}
\label{sec:review}

Let us begin with a precise statement of the problem. We consider an
infrared safe three-jet-like observable in $e^+e^- \to {\it hadrons}$
such as a particular moment of the thrust distribution. The observable
can be expanded in powers of
$\alpha_s/\pi$,
\begin{equation}
\sigma = \sum_n 
\sigma^{[n]},
\hskip 1 cm
\sigma^{[n]} \propto \left(\alpha_s / \pi\right)^n\,.
\end{equation}
The order $\alpha_s^2$ contribution has the form
\begin{eqnarray}
\sigma^{[2]} &=&
{1 \over 2!}
\int d\vec p_1 d\vec p_2\
{d \sigma^{[2]}_2 \over d\vec p_1 d\vec p_2}\
{\cal S}_2(\vec p_1,\vec p_2)
\nonumber\\
&&+
{1 \over 3!}
\int d\vec p_1 d\vec p_2 d\vec p_3\
{d \sigma^{[2]}_3 \over d\vec p_1 d\vec p_2 d\vec p_3}\
{\cal S}_3(\vec p_1,\vec p_2,\vec p_3)
\label{start}\\
&&
+
{1 \over 4!}
\int d\vec p_1 d\vec p_2 d\vec p_3 d\vec p_4\
{d \sigma^{[2]}_4 \over d\vec p_1 d\vec p_2 d\vec p_3 d\vec p_4}\
{\cal S}_4(\vec p_1,\vec p_2,\vec p_3,\vec p_4).
\nonumber
\end{eqnarray}
Here the $d\sigma^{[2]}_n$ are the  order $\alpha_s^2$ contributions
to the parton level cross section, calculated with zero quark masses.
Each contains momentum and energy conserving delta functions. The $d
\sigma^{[2]}_n$ include ultraviolet renormalization in the
$\overline{\rm MS}$ scheme. The functions $\cal S$ describe the
measurable quantity to be calculated. We wish to calculate a
``three-jet-like'' quantity.  That is, ${\cal S}_2 = 0$. The
normalization is such that ${\cal S}_n = 1$ for $n = 2,3,4$ would give
the order $\alpha_s^2$ perturbative contribution the total cross
section.  There are, of course, infrared divergences associated with
Eq.~(\ref{start}). For now, we may simply suppose that an infrared
cutoff has been supplied.

The measurement, as specified by the functions ${\cal S}_n$, is to be
infrared safe, as described in Ref.~\cite{KS}: the ${\cal S}_n$ are
smooth functions of the parton momenta and
\begin{equation}
{\cal S}_{n+1}(\vec p_1,\dots,\lambda \vec p_n,(1-\lambda)\vec p_n)
= 
{\cal S}_{n}(\vec p_1,\dots, \vec p_n)
\end{equation}
for $0\le \lambda <1$. That is, collinear splittings and soft
particles do not affect the measurement.

It is convenient to calculate a quantity that is dimensionless. Let the
functions ${\cal S}_n$ be dimensionless and eliminate the remaining
dimensionality in the problem by dividing by $\sigma_0$, the total
$e^+ e^-$ cross section at the Born level. Let us also remove the
factor of $(\alpha_s / \pi)^2$. Thus, we calculate
\begin{equation}
{\cal I} = {\sigma^{[2]} \over \sigma_0\ (\alpha_s/\pi)^2}.
\label{calIdef}
\end{equation}

Let us now see how to  set up the calculation of ${\cal I}$ in a
convenient form. We note that ${\cal I}$ is a function of the c.m.\
energy $\sqrt s$ and the $\overline{\rm MS}$ renormalization scale
$\mu$. We will choose $\mu$ to be proportional to $\sqrt s$: $\mu =
A_{UV} \sqrt s$. Then ${\cal I}$ depends on $A$. But, because it is
dimensionless, it is independent of $\sqrt s$. This allows us to write
\begin{equation}
{\cal I} = \int_0^\infty d \sqrt s\ h(\sqrt s)\ 
{\cal I}(A_{UV},\sqrt s),
\end{equation}
where $h$ is any function with
\begin{equation}
\int_0^\infty d \sqrt s\ h(\sqrt s) = 1.
\label{rtsintegral}
\end{equation}

\begin{figure}
\centerline{\DESepsf(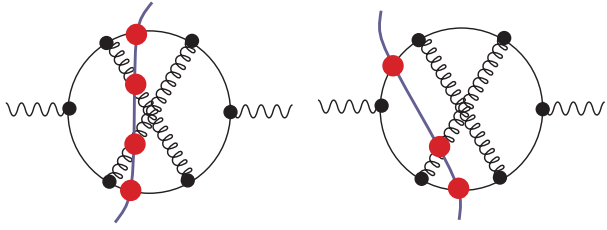 width 8 cm)}
\medskip
\caption{Two cuts of one of the Feynman diagrams that contribute to 
$e^+e^- \to {\it hadrons}$.}
\label{fig:cutdiagrams}
\end{figure}

The quantity $\cal I$ can be expressed in terms of cut Feynman
diagrams, as in Fig.~\ref{fig:cutdiagrams}. The dots where the parton
lines cross the cut represent the function ${\cal S}_n(\vec p_1,
\dots ,\vec p_n)$. Each diagram is a three loop diagram, so we have
integrations over loop momenta $ l_1^\mu$, $ l_2^\mu$ and
$ l_3^\mu$. We first perform the energy integrations. For the graphs
in which four parton lines cross the cut, there are four mass-shell
delta functions $\delta (k_J^2)$. These delta functions eliminate the
three energy integrals over $ l_1^0$, $ l_2^0$, and $ l_3^0$ as
well as the integral (\ref{rtsintegral}) over $\sqrt s$. For the
graphs in which three parton lines cross the cut, we can eliminate the
integration over $\sqrt s$ and two of the $ l_J^0$ integrals. One
integral over the energy $E$ in the virtual loop remains. We perform
this integration by closing the integration contour in the lower half
$E$ plane. This gives a sum of terms obtained from the original
integrand by some simple algebraic substitutions. Having performed the
energy integrations, we are left with an integral of the form
\begin{equation}
{\cal I} = 
\sum_G
\int d\vec  l_1\,d\vec  l_2\,d\vec  l_3\
\sum_C\,
g(G,C;\vec l_1,\vec l_2,\vec l_3).
\label{master}
\end{equation}
Here there is a sum over graphs $G$ (of which one is shown in
Fig.~\ref{fig:cutdiagrams}) and there is a sum over the possible cuts
$C$ of a given graph. The problem of calculating ${\cal I}$ is now set
up in a convenient form for calculation.

If we were using the Ellis-Ross-Terrano method, we would put the sum
over cuts outside of the integrals in Eq.~(\ref{master}).
For those cuts $C$ that  have three partons in the final state, there
is a virtual loop. We can arrange that one of the loop momenta, say
$\vec l_1$, goes around this virtual loop. The essence of the ERT
method is to perform the integration over the virtual loop momentum
analytically ahead of time. The integration is often ultraviolet
divergent, but the ultraviolet divergence is easily removed by a
renormalization subtraction. The integration is also typically infrared
divergent. This divergence is regulated by working in $3 - 2\epsilon$
space dimensions and then taking $\epsilon \to 0$ while dropping the
$1/\epsilon^n$ contributions (after proving that they cancel against
other contributions). After the $\vec l_1$ integration has been
performed analytically, the integrations over $\vec  l_2$ and $\vec
 l_3$ can be performed numerically. For the cuts $C$ that have four
partons in the final state, there are also infrared divergences. One
uses either a ``phase space slicing'' or a ``subtraction'' procedure to
get rid of these divergences, cancelling the $1/\epsilon^n$ pieces
against the $1/\epsilon^n$ pieces from the virtual graphs. In the end,
we are left with an integral $\int d\vec  l_1\,d\vec  l_2\,d\vec  l_3$
in exactly three space dimensions that can be performed numerically.

In the numerical method, we keep the sum over cuts $C$ inside the
integrations. We take care of the ultraviolet divergences by simple
renormalization subtractions on the integrand. We make certain
deformations on the integration contours so as to keep away from poles
of the form $1/[E_F - E_I \pm i\epsilon]$, where $E_F$ is the energy of
the final state and $E_I$ is the energy of an intermediate state. Then
the integrals are all convergent and we calculate them by Monte Carlo
numerical integration.

Let us now look at the contour deformation in a little more detail. We
denote the momenta $\{\vec l_1,\vec l_2,\vec l_3\}$ collectively
by $ l$ whenever we do not need a more detailed description. Thus
\begin{equation}
{\cal I} = 
\sum_G
\int\! d l\
\sum_C\,
g(G,C; l).
\label{basicagain}
\end{equation}
For cuts $C$ that leave a virtual loop integration, there are
singularities in the integrand of the form $E_F - E_I +i\epsilon$ (or
$E_F - E_I -i\epsilon$ if the loop is in the complex conjugate
amplitude to the right of the cut). Here $E_F$ is the energy of the
final state defined by the cut $C$ and $E_I$ is the energy of a
possible intermediate state. We will examine the nature of these
``scattering singularities'' in some detail in the next section. For
now, all we need to know is that $E_F - E_I = 0$ on a surface in the
space of $\vec l_1$ for fixed $\vec l_2$ and $\vec l_3$ if we
pick $\vec l_1$ to be the momentum that flows around the virtual
loop. These singularities do not create divergences. The Feynman rules
provide us with the $i\epsilon$ prescriptions that tell us what to do
about the singularities: we should deform the integration contour into
the complex $\vec l_1$ space so as to keep away from them. Thus we
write our integral in the form
\begin{equation}
{\cal I} = 
\sum_G
\int\! d l\
\sum_C\,
{\cal J}(G,C; l)\,
g(G,C; l+i\kappa(G,C; l)).
\label{deformed}
\end{equation}
Here $i\kappa$ is a purely imaginary nine-dimensional vector that we add
to the real nine-dimensional vector $ l$ to make a complex
nine-dimensional vector. The imaginary part $\kappa$ depends on the
real part $ l$, so that when we integrate over $ l$, the
complex vector $ l + i\kappa$ lies on a surface, the integration
contour, that is moved away from the real subspace. When we thus deform
the contour, we supply a jacobian ${\cal J} = \det(\partial ( l +
i\kappa)/\partial  l)$. (See Ref.~\cite{beowulfPRD} for details.) 

The amount of deformation $\kappa$ depends on the graph $G$ and,
more significantly, the cut $C$. For cuts $C$ that leave no virtual
loop, each of the momenta $\vec  l_1$, $\vec  l_2$, and $\vec
 l_3$ flows through the final state. For practical reasons, we want
the final state momenta to be real. Thus we set $\kappa = 0$ for cuts
$C$ that leave no virtual loop. On the other hand, when the cut $C$
does leave a virtual loop, we choose a non-zero $\kappa$. We must,
however, be careful. When $\kappa = 0$ there are singularities in $g$
on certain surfaces that correspond to collinear parton momenta. These
singularities cancel between $g$ for one cut $C$ and $g$ for another.
This cancellation would be destroyed if, for $ l$ approaching the
collinear singularity, $\kappa = 0$ for one of these cuts but not for
the other. For this reason, we insist that for all cuts $C$, $\kappa \to
0$ as $ l$ approaches one of the collinear singularities. The details
can be found in Ref.~\cite{beowulfPRD}. All that is important here is
that $\kappa \to 0$ quadratically with the distance to a collinear
singularity.

Much has been left out in this brief overview, but we should now have
enough background to see the requirements for a sensible choice of
integration points.

\section{General principles for the choice of points}
\label{sec:general}

In the following sections, we discuss the choice of integration points
for the evaluation of a given graph. In this section, we summarize
the general principles of Monte Carlo integration as they apply to
this calculation.

We wish to perform an integral of the form
\begin{equation}
{\cal I} = \sum_G\int\! d l\ f(G; l)
\end{equation}
where 
\begin{equation}
f(G; l) =
\Re \left\{
\sum_C\,
{\cal J}(G,C; l)\,
g(G,C; l+i\kappa(G,C; l))
\right\}.
\end{equation}
(We can take the real part because we know in advance that $\cal I$ is
real.) Using the Monte Carlo technique, for each graph $G$ we choose a
large number $\xi(G) N$ of points $ l$, with $\sum_G\xi(G) = 1$ so
that the total number of points is $N$. We sample the points $ l$ for
graph $G$ at random with a density $\rho(G; l)$, with normalization
$\int\! d l\,\rho(G; l) = 1$. Then the integral is approximated by 
\begin{equation}
{\cal I}_N = { 1 \over N}\sum_G { 1 \over \xi(G)}
\sum_{j = 1}^{\xi(G)N} { f( l_j) \over \rho(G; l_j)}.
\end{equation}

This is an approximation for the integral in the sense that, if we
repeat the procedure a lot of times, the expectation value for 
${\cal I}_N$ is 
\begin{equation}
\langle {\cal I}_N \rangle = {\cal I}.
\end{equation}
The expected r.m.s.\ error is ${\cal E}$, where
\begin{equation}
{\cal E}^2 = \langle \left( {\cal I}_N - {\cal I}\right)^2\rangle.
\end{equation}
With a little manipulation, one can rewrite this as
\begin{equation}
{\cal E}^2 =
{ 1 \over N}\left\{ \sum_G\, \xi(G)
\left[
{ \Delta(G) \over \xi(G)}
- \sum_L \Delta(L) 
\right]^2
+ \left(\sum_L \Delta(L) \right)^2
\right\}.
\end{equation}
where
\begin{eqnarray}
\Delta(G)^2 &=&
\int d l\ \rho(G; l)
\left[
{ |f(G; l)| \over \rho(G; l)}
- \int d l'\ |f(G; l')|
\right]^2
\nonumber\\
&&+
\int d l\ \Bigl(|f(G; l)| + f(G; l)\Bigr)
\times
\int d l'\ \Bigl(|f(G; l')| - f(G; l')\Bigr).
\end{eqnarray}
We see, first of all, that the expected error decreases proportionally
to $1/\sqrt N$. Second, we see that for a given choice of the density
functions $\rho(G; l)$, the ideal choice of the factors $\xi(G)$ is
$\xi(G) \propto \Delta(G)$. This is, in fact, easy to implement. Third,
the ideal choice of $\rho(G; l)$ is $\rho(G; l) \propto |f(G; l)|$.
This is, in fact, impossible to implement. 

Although it is not possible to choose $\rho(G; l) \propto
|f(G; l)|$, at least we can choose it so that ${|f(G; l)| /
\rho(G; l)}$ is not singular at the singularities of $|f( l)|$.
Furthermore, we can try to make $\rho(G; l)$ big at places where we
know that $|f(G; l)|$ is big.

We will build the general sampling method out of elementary sampling
methods. That is, we will find a number of simple methods to choose
points $ l$ for our graph. Let the density of points for
the $i$th elementary sampling method be $\rho_i(G; l)$  (normalized
to $\int\!d l\,\rho_i(G; l) = 1$). Then we devote a fraction
$\lambda_i(G)$ of the points to the choice with density $\rho_i(G; l)$
and obtain a net density
\begin{equation}
\rho(G; l) = \sum_i \lambda_i(G)\, \rho_i(G; l).
\end{equation}
In this way, we make the sampling problem manageable. If we know that
$|f(G; l)|$ is big near a certain point or surface in the space of
loop momenta, we can design one of the elementary sampling methods so
that the corresponding $\rho_i(G; l)$ is big there. In undertaking
this task, we do not have to simultaneously arrange that 
$\rho_i(G; l)$ be big at other places where  $|f(G; l)|$ is big.

In the following sections, we discuss the elementary sampling methods. 
We imagine that we are dealing with only one specific graph $G$, so we
suppress the index $G$ in the notation.

\section{Organization of the sampling method}
\label{sec:organization}

We consider a fixed (uncut) graph, dropping references to the label $G$
of the graph. As mentioned in the previous section, we sample points
$ l$ in the space of loop momenta according to several elementary
sampling methods, each labelled by and index $i$ and having a density
of points $\rho_i( l)$. The net density is then $\rho( l) =
\sum \lambda_i  \rho_i( l)$.

We first address the identification of loop momenta. There are eight
propagator momenta $\vec k_P$. (See, for example,
Fig.~\ref{fig:cutdiagrams}.) The three loop momenta
$\vec  l_L$ can in general be any three linearly independent linear
combinations of the $\vec k_P$. We will keep the choice simple by
choosing three of the  $\vec k_P$ to be the loop momenta. However, this
still leaves us with the choice  of which three of the $\vec k_P$
should be loop momenta. It proves convenient to make {\em different}
choices for different elementary sampling methods. We specify the
choice by specifying a triplet of integers $\{Q(1),Q(2),Q(3)\}$ such
that $\vec l_J$ is  $\vec k_{Q(J)}$. We call $Q$ an index set. Then the
complete set of propagator matrices can related to the loop momenta by
a matrix $A$:
\begin{equation}
\vec k_P = \sum_{L=1}^3 A^P_L \vec l_L.
\end{equation}
Evidently, given the index set $Q$, the matrix $A$ can be constructed
by using the topology of the graph.

Now we characterize certain surfaces, to be called scattering
singularity surfaces, near which the integrand $|f( l)|$ is big. To
do this, we consider the cuts of our graph in which three partons
appear in the final state. For each such cut, there is a virtual loop.
Let $\vec  l_1$ be the loop momentum. More precisely,
$\vec  l_1$ will be the momentum, $\vec k_{Q(1)}$, of one of the
propagators in the loop, but we defer for a moment specifying which
one. As specified in Sec.~\ref{sec:review}, the integration contour for
$\vec  l_1$ is deformed into the complex plane,\footnote{We keep the
momenta that enter the final state real.} so that $\vec  l_1
\to \vec  l_{1,c} = \vec  l_1 + i\vec \kappa$.

\begin{figure}
\centerline{\DESepsf(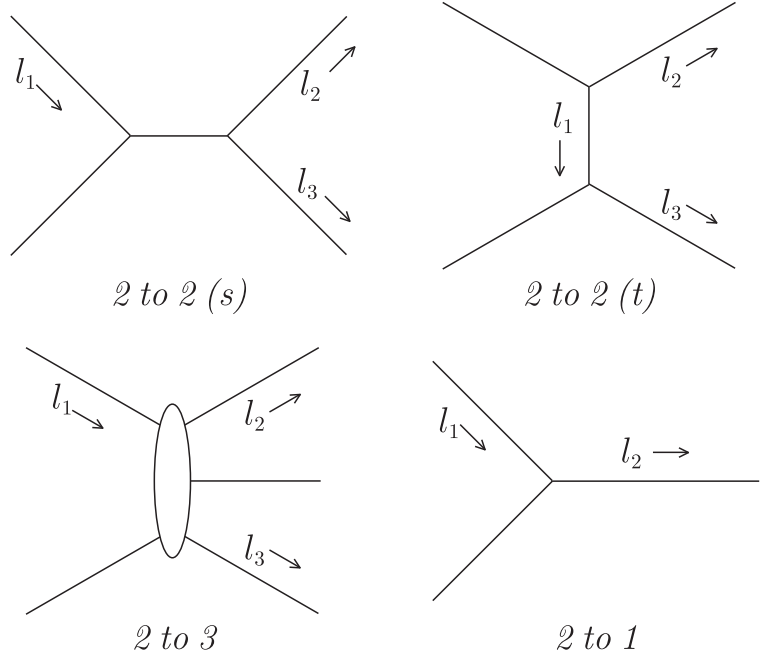 width 10 cm)}
\medskip
\caption{Elementary scattering subdiagrams that occur at
next-to-leading order.}
\label{fig:scatterings}
\end{figure}

\begin{figure}
\centerline{\DESepsf(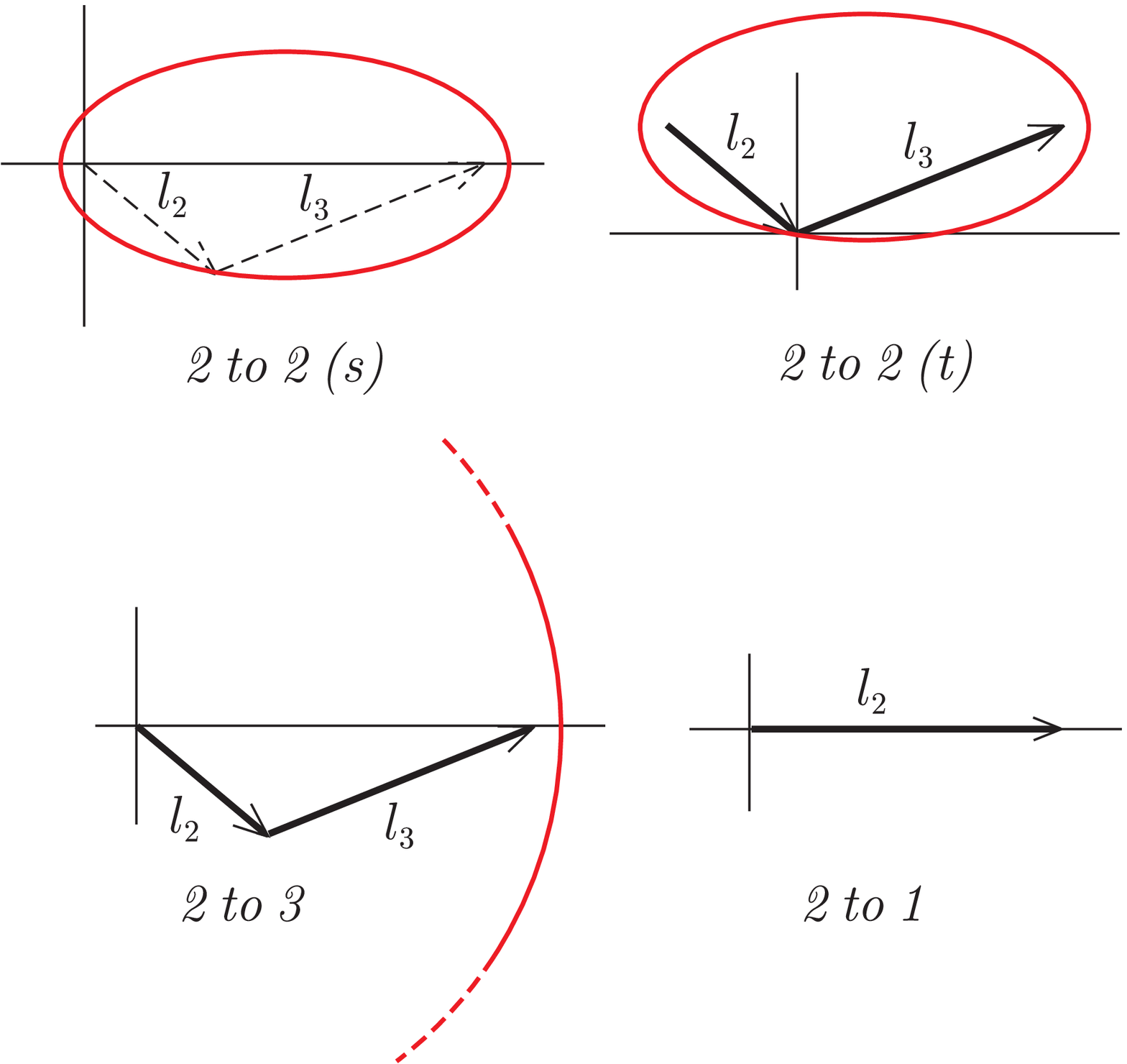 width 10 cm)}
\medskip
\caption{Singularity surfaces associated with the elementary
scatterings in Fig.~\protect\ref{fig:scatterings}. In each case,
the vectors $\vec l_2$ and $\vec l_3$ (or just $\vec l_2$ for {\it 2 to
1} scattering) are indicated by arrows. We see the scattering
singularity surface in the space of  $\vec l_1$. For {\it 2 to 2 (t)}
scattering and {\it 2 to 2 (s)} scattering, these surfaces are
ellipsoids. For {\it 2 to 3} scattering, the surface is a sphere, only
part of which is shown. For {\it 2 to 1} scattering, the surface reduces
to a line segment. The integrand is typically not singular on the
scattering singularity surface because of the contour deformation.
However, the contour deformation vanishes along the heavy straight
lines. Thus, in particular, in the {\it 2 to 2 (t)} case the integrand
is actually singular at $\vec l_1 = 0$.}
\label{fig:singularities}
\end{figure}

Before deformation,
the integrand is singular on certain surfaces associated with simple
scattering processes, the scattering singularity surfaces. How can
this happen? There are four cases. Each case is illustrated by a
Feynman graph in Fig.~\ref{fig:scatterings}. The corresponding 
scattering singularity surface is illustrated in
Fig.~\ref{fig:singularities}. The cases are

\begin{enumerate}

\item {\it 2 to 2 (s)}. 
A virtual parton with momentum $\vec  l_1$ merges with a virtual parton
with momentum $\vec  l_2 + \vec  l_3 - \vec  l_1$ to produce a virtual
parton with momentum $\vec l_2 + \vec  l_3$. This virtual parton divides
into two partons with  momenta $\vec  l_2$ and $\vec  l_3$ that enter
the final state.  In old-fashioned perturbation theory, there is an
energy denominator $|\vec  l_2|+|\vec  l_3| - |\vec  l_1| - |\vec  l_2
+ \vec  l_3 - \vec  l_1| + i\epsilon$, which vanishes on the ellipsoid
$| \vec  l_1| + |\vec  l_2 + \vec  l_3 - \vec  l_1| = |\vec l_2|+|\vec 
l_3|$ in the space of $\vec  l_1$.  This is the scattering singularity
surface. The contour deformation is non-zero on the entire scattering
singularity surface.

\item {\it 2 to 2 (t)}.
A virtual parton with momentum $\vec  l_2 + \vec  l_1$ scatters from a
virtual parton with momentum $\vec  l_3 - \vec l_1$ by exchanging a
parton with momentum $\vec  l_1$. Partons with momentum $\vec  l_2$ and
$\vec  l_3$ emerge into the final state.  There is a scattering
singularity surface $|\vec  l_2 + \vec  l_1| + |\vec  l_3 - \vec  l_1| =
|\vec  l_2|+|\vec  l_3|$. In this case, there is also a singularity at
$\vec  l_1 = 0$ that arises from the propagator of the exchanged
parton. This soft exchange singularity lies on the scattering
singularity surface. Furthermore, in our treatment, the contour
deformation vanishes at the soft exchange singularity so that, even
after contour deformation, the integrand is singular there. This is,
however, an integrable singularity.

\item {\it 2 to 3}.
A virtual parton with momentum $\vec  l_1$ collides with a
virtual parton with momentum $-\vec  l_1$ to produce a final state
with partons carrying momenta $\vec  l_2$, $\vec  l_3$, and $-\vec
l_2 - \vec  l_3$. There is an energy denominator $|\vec  l_2|
+ |\vec  l_3| + |\vec  l_2 + \vec  l_3| - 2 |\vec  l_1| +
i\epsilon$, which is singular on the sphere $|\vec  l_1| =
[|\vec  l_2| + |\vec  l_3| + |\vec  l_2 +\vec  l_3|]/2$.
The contour deformation is non-zero on the entire scattering singularity
surface.

\item {\it 2 to 1}.
A virtual parton with momentum $\vec  l_1$ combines with a
virtual parton with momentum $\vec  l_2 - \vec  l_1$ to produce an
on shell parton with momentum $\vec  l_2$ that enters the final state.
There is an energy denominator $|\vec  l_2| - |\vec  l_1| -
|\vec l_2 - \vec l_1| + i\epsilon$, which vanishes on the line
$\vec  l_1 = x \vec  l_2$ with $0\le x \le 1$. The contour
deformation is chosen to vanish at this collinear singularity. As
discussed in Ref.~\cite{beowulfPRD}, the collinear singularity cancels
when one sums over cuts. However, singularities at $\vec  l_1 = 0$ an
$\vec  l_1 = \vec  l_2$ remain.

\end{enumerate} 

Thus for each cut there are a number of scattering singularity
surfaces. There is a contour deformation that keeps the integrand from
being singular except at special points on these surfaces. However, the
integrand is sometimes still quite large near these surfaces. For this
reason, we will design an elementary sampling method for each surface
such that the density of points is big on the whole surface and
singular at the special point if necessary. 

There are two kinds of singularities associated with points $\vec k_P =
0$ where a propagator momentum vanishes. First, there is a 
$1/|\vec k_P|^2$ singularity when the momentum of the exchanged parton
in a {\it 2 to 2 (t)} scattering vanishes. The elementary sampling
method associated with the {\it 2 to 2 (t)} scattering will be designed
to take care of this kind of singularity. Second, there is a $1/|\vec
k_P|$ singularity when {\it any} propagator momentum $\vec k_P$
approaches zero. This weak singularity arises simply because massless
Lorentz invariant phase space is $d\vec k/|\vec k|$. As it turns out,
these singularities in the density will automatically be provided by
the combined sampling methods without having to specifically arrange
for them.

The {\it 2 to 1} scattering singularity surface is exceptional in the
list above in that there is no singularity except for the two singular
points at $\vec l_1 = 0$ and $\vec l_2 - \vec l_1 = 0$. These two
singular points are of one or the other of two types
mentioned above. Typically a {\it 2 to 1} scattering subdiagram is part
of a {\it 2 to 2 (s)} or {\it 2 to 2 (t)} scattering subdiagram and the
two singularities are provided for by the {\it 2 to 2
(s)} or {\it 2 to 2 (t)} sampling methods. The
exception is in the case of a self-energy subgraph that is connected to
a final state parton. In this case, the {\it 2 to 2 (t)}, {\it 2 to 2
(s)}, and {\it 2 to 3} sampling methods do not apply and we need an
explicit {\it 2 to 1} sampling method. Thus we will apply a 
{\it 2 to 1} sampling method only in the case of a self-energy subgraph
connected to a final state parton.

The previous argument indicates that for each scattering singularity
surface in the space of the loop momentum $\vec  l_1$ in a virtual
loop, we should associate a method for choosing $\vec  l_1$ that puts a
high density of points near this surface.  It is then useful to choose
the other two loop momenta to be the momenta of two of the three
partons in the final state. Thus the momenta of the final state partons
are $\vec  l_2$, $\vec  l_3$, and $-\vec  l_2 - \vec  l_3$. The
integrand will be singular whenever the three partons approach a two jet
configuration. Thus we choose the points $\{\vec  l_2,\vec  l_3\}$ so
that the density of points is appropriately singular at the two jet
configurations.

\begin{figure}
\centerline{\DESepsf(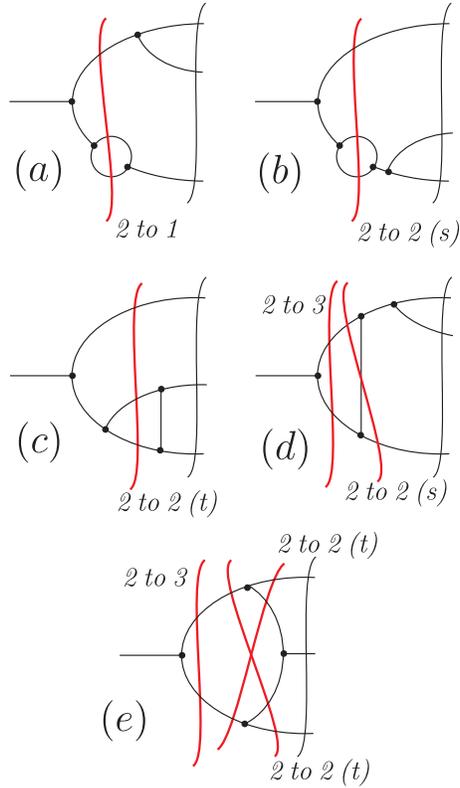 width 6 cm)}
\medskip
\caption{Matching of scattering singularities to the structure of one
loop virtual subdiagrams. A scattering singularity occurs when the
energy of an intermediate state matches the energy of the final state.
For each diagram, the relevant intermediate states are marked with a
line through the graph. The label near the line indicates the type of
the corresponding singularity. For some graphs, there is more than one
scattering singularity, as indicated. The {\it 2 to 1} singularity is
marked only in the case of a self-energy subdiagram connected to a
final state parton.}
\label{fig:virtualtypes}
\end{figure}

Thus we have a general scheme for organizing the elementary
sampling methods. First, find all of the possible three  parton cuts of
the graph in question. Then, for each such final state cut,
enumerate the scattering singularity surfaces that can occur, counting
the {\it 2 to 1} case only when the virtual loop is a self-energy
diagram connected to a final state parton. There are five basic
situations, as illustrated in Fig.~\ref{fig:virtualtypes}. Each
combination of a final state cut and a scattering singularity surface
will correspond to an elementary sampling method.

One point should be emphasized for clarity. In the numerical method
discussed in this paper, for each integration point, we compute
$f/\rho$ where $f$ is the integrand and $\rho$ is the density of
integration points. Contributions from all cuts of a given graph are
calculated and summed to form the integrand $f$. The density $\rho$ is
also a sum, with terms corresponding to each of the possible cuts of
the graph. Thus there are {\it independent} sums over possible cuts in
$f$ and in $\rho$.

Consider the five basic amplitudes with a virtual loop that are
illustrated in Fig.~\ref{fig:virtualtypes}. In two of these cases,
there are more than one possible intermediate state involving the
virtual loop and thus more than one scattering singularity surface. In
these cases, it is important to understand how the scattering
singularity surfaces fit together.

\begin{figure}
\centerline{\DESepsf(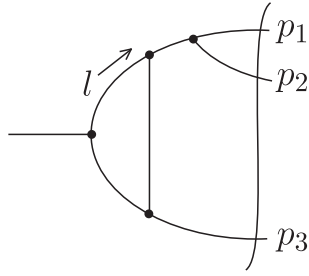 width 4 cm)}
\medskip
\caption{Labelling of momenta for graphs of type (d) in
Fig.~\protect\ref{fig:virtualtypes}.}
\label{fig:virtuald}
\end{figure}

\begin{figure}
\centerline{\DESepsf(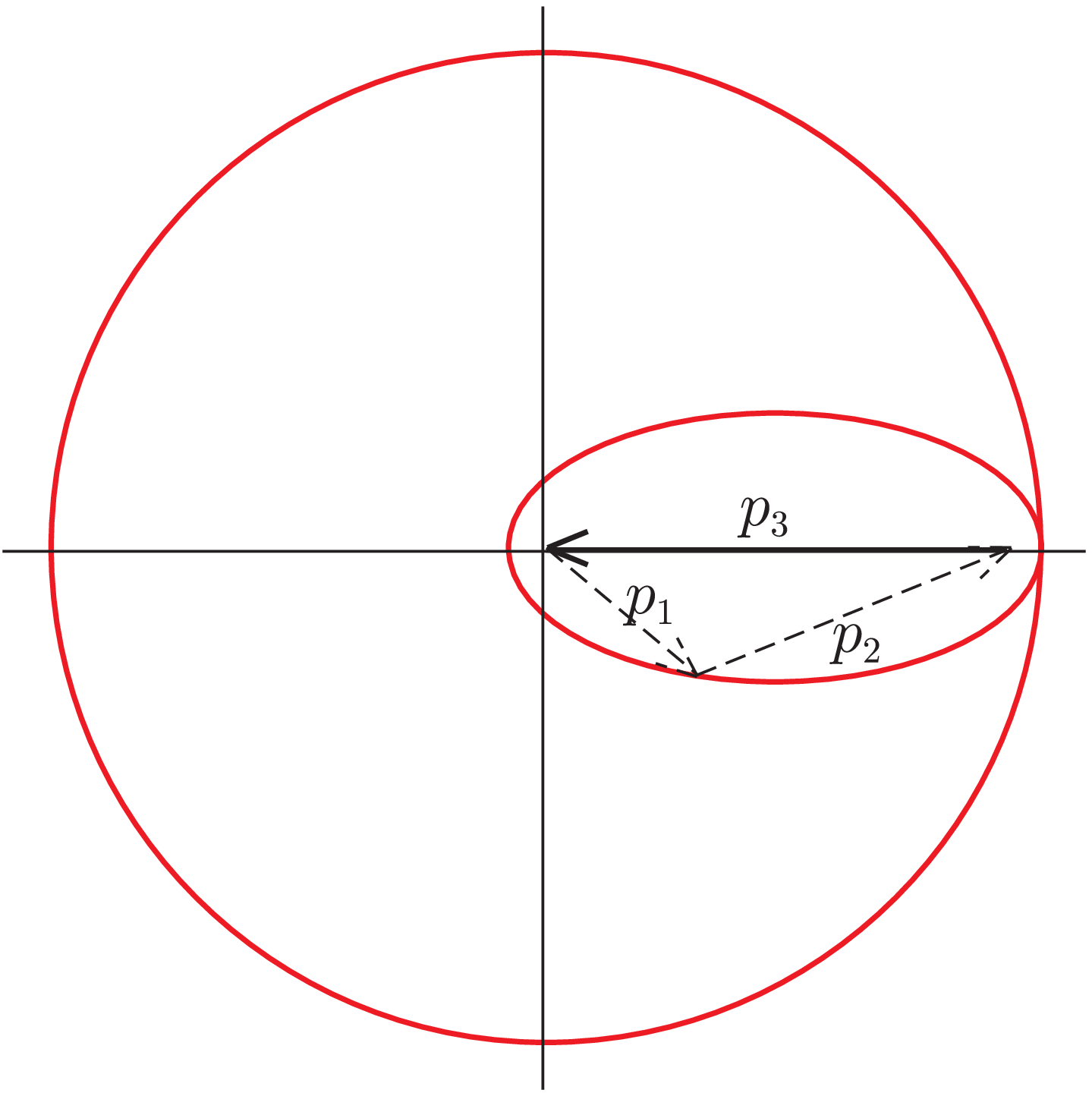 width 8 cm)}
\medskip
\caption{Singularity surfaces for graphs of type (d) in
Fig.~\protect\ref{fig:virtualtypes}.}
\label{fig:singsd}
\end{figure}

Let us examine, then, case $(d)$ in Fig.~\ref{fig:virtualtypes}. Label
the momenta as in Fig.~\ref{fig:virtuald}. In Fig.~\ref{fig:singsd} we
see the scattering singularity surfaces in the space of the loop
momentum $\vec l \equiv \vec l_1$ for an arbitrary fixed choice of the
final state momenta $\vec p_j$. (We draw the figure for $\vec l$ in the
plane of the $\vec p_j$). The vectors $\vec p_1$, $\vec p_2$, and
$\vec p_3$ are indicated. The integration contour is deformed
everywhere except along the line $\vec l = - x \vec p_3$ with $0<x<1$,
which is indicated as a solid line. There is an ellipsoidal {\it 2 to
2 (s)} surface, $|\vec l| + |\vec p_1 + \vec p_2 - \vec l| =  |\vec
p_1| + |\vec p_2|$. There is also a spherical {\it 3 to 2} surface,
$2|\vec l| = |\vec p_1| + |\vec p_2| + |\vec p_3|$. The integration
contour is deformed everywhere on both of these surfaces, so that the
integrand is not singular anywhere. On the other hand, the deformation
vanishes on the solid line, which can be rather near the ellipsoidal
surface in the case that the angle between $\vec p_1$ and $\vec p_2$
is small. Thus the integrand can be rather big on the ellipsoid in
this circumstance. Furthermore, the size of the integrand is enhanced
where the ellipsoid is tangent to the spherical singularity surface.
We will need to put an extra density of points in the region of this
point of tangency.

\begin{figure}
\centerline{\DESepsf(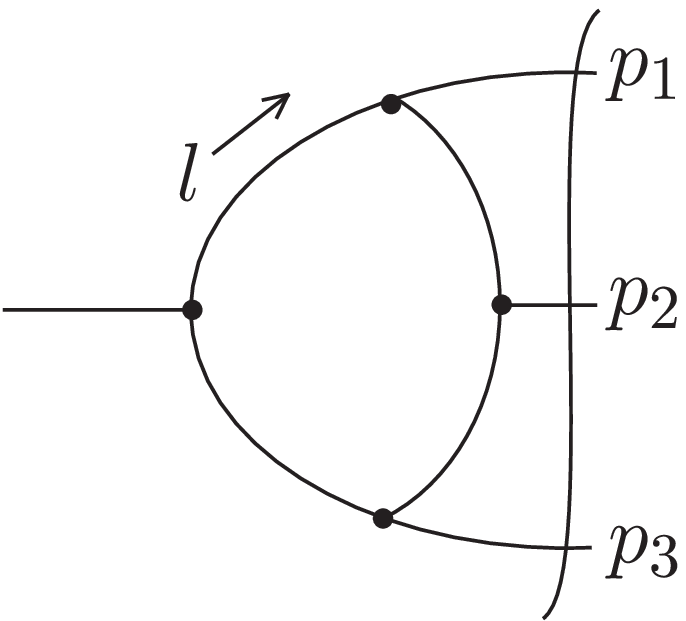 width 4 cm)}
\medskip
\caption{Labelling of momenta for graphs of type (e) in
Fig.~\protect\ref{fig:virtualtypes}.}
\label{fig:virtuale}
\end{figure}

Case $(e)$ in Fig.~\ref{fig:virtualtypes} just a little more
complicated. Label the momenta as in Fig.~\ref{fig:virtuale}.
Fig.~\ref{fig:singsd} shows the scattering singularity surfaces in the
space of the loop momentum $\vec l$ for fixed final state momenta $\vec
p_j$. The vectors $\vec p_1$, $\vec p_2$, and $\vec p_3$ are indicated.
The integration contour is deformed everywhere except along the lines
$\vec l =  x \vec p_1$, $\vec l - \vec p_1=  x \vec p_2$ and  $\vec l
- \vec p_1 - \vec p_2 =  x \vec p_3$, with $0<x<1$ in each case. These
lines are are indicated as solid lines that form a triangle. There
are two ellipsoidal {\it 2 to 2 (t)} surfaces, $|\vec l| + |\vec p_1
+ \vec p_2 - \vec l| =  |\vec p_1| + |\vec p_2|$ and $|\vec l - \vec
p_1| + |\vec l| =  |\vec p_2| + |\vec p_3|$. There is also a spherical
{\it 3 to 2} surface, $2|\vec l| = |\vec p_1| + |\vec p_2| + |\vec
p_3|$.  As in the previous case, $(d)$, we need an especially high
density of integration points near where an ellipsoid is tangent to the
sphere in the case that this point is near to the triangle, where the
deformation vanishes. We have also the new feature that each of the
ellipsoids shares a point with the triangle. At this point the contour
deformation is zero, so the integrand is actually singular. This is the
point where the momentum of the exchanged parton in the associated {\it
2 to 2 (t)} scattering vanishes. The density of integration points will
have to have a corresponding singularity at these points.

\begin{figure}
\centerline{\DESepsf(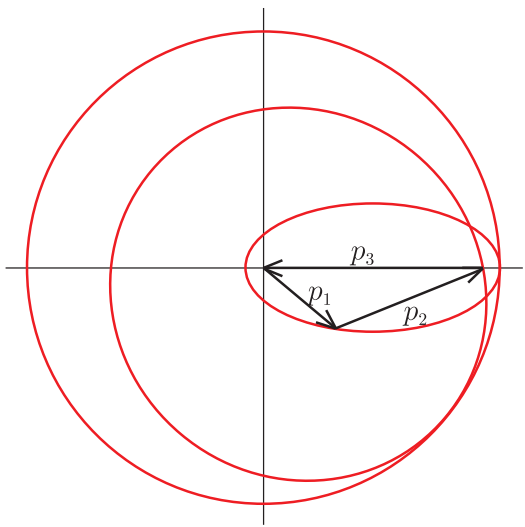 width 8 cm)}
\medskip
\caption{Singularity surfaces for graphs of type (e) in
Fig.~\protect\ref{fig:virtualtypes}.}
\label{fig:singse}
\end{figure}

This completes the description of the general organization of the
sampling scheme. In the following sections, we outline its component
parts.

\section{Sampling for the final state}
\label{sec:finalstate}

As discussed in the previous section, for each final state cut with
three partons in the final state, we will define a number of elementary
sampling methods. Part of each method is to choose with an
appropriate density the momentum $\vec l_1$ that circles through the
virtual loop that occurs when there are three final state partons. The
other part is to choose the momenta $\vec  l_2$, $\vec  l_3$, and
$-\vec  l_2 - \vec  l_3$ of the final state partons. We address the
choice of the final state momenta in this section.

Let the final state parton momenta be $\vec p_1$, $\vec p_2$, and $\vec
p_3$, with  $\sum_i \vec p_i = 0$. (Two of these will then be $\vec
 l_2$ and $\vec  l_3$, but it doesn't matter which ones.) Define
$E_{\rm max} = {\scriptstyle {1\over 2}}\sum_i |\vec p_i|$ and $x_i =
|\vec p_i|/E_{\rm max}$. Then $\sum_i x_i = 2$ and $0 \le x_i \le 1$. We
can write the integration over final state momenta as follows, using
$p_3^2 = p_1^2 + p_2^2 + 2 p_1 p_2 \cos\theta_{12}$ so that
$d\cos\theta_{12} = p_3 dp_3/(p_1 p_2) $:
\begin{eqnarray}
{\cal I}&=&\int d\vec p_1\,d\vec p_2\,d\vec p_3\ \delta(\sum\vec
p_i)\,f(p)
\nonumber\\
&=& \int 2 E_{\rm max}^6 d\ln E_{\rm max}\, 
d\cos\theta_1\, d\phi_1\,d\phi_{12}\,
x_1 x_2 x_3\, dx_1\,  dx_2\, 
f(p).
\end{eqnarray}
Here $\{\theta_1,\phi_1\}$ are the angles of $\vec p_1$ and $\phi_{12}$
is the angle between $\vec p_1 \times \vec p_2$ and $\vec p_1 \times
\vec e$, where $\vec e$ is an arbitrary reference vector.

If we sample $\{ \ln E_{\rm max}, \cos\theta_1, \phi_1, \phi_{12}\}$
with a density
\begin{equation}
\rho_A \equiv 
{ dN \over d\ln E_{\rm max}\, 
d\cos\theta_1\, d\phi_1\,d\phi_{12}}
\end{equation}
and we sample $\{x_1, x_2\}$ with a density
\begin{equation}
\rho_B \equiv 
{ dN \over dx_1\,dx_2},
\end{equation}
then the net density is
\begin{equation}
\rho \equiv { dN \over d\vec p_1\,d\vec p_2}
= { \rho_A \rho_B \over 2 E_{\rm max}^6\, x_1 x_2 x_3}.
\label{rhonet}
\end{equation}
On symmetry grounds, $\rho_A$ should be independent of the angles. Its
dependence on $E_{\rm max}$ is not too important. A convenient choice,
normalized so that $\int dN = 1$, is
\begin{equation}
{1\over\rho_A} = {8 \pi^2 \over 3}\,\left({E_0/ E_{\rm max}}\right)^3
\left[1 +  \left({E_{\rm max}/ E_0}\right)^3\right]^2,
\end{equation}
where $E_0$ is a fixed parameter with dimensions of mass.

What should be the properties of $\rho_B$? We can expect the integrand
$f(p)$ to be singular in any two jet region: when any of the $\vec p_j$
vanishes ($x_j \to 0$) or when two of them become collinear (if $\vec
p_1$ and $\vec p_2$ are nearly collinear, then $x_3 \approx 1$). The 
factor $1/x_1 x_2 x_3$ in Eq.~(\ref{rhonet}) provides a weak
singularity in the $x_i \to 0$ regions. We can build in a weak
singularity in the collinear regions by choosing
\begin{equation}
\rho_B = 
{ {\cal N}_x\, \max_j\sqrt{1-x_j} \over \prod_j \sqrt{1-x_j}}.
\end{equation}
The $x_i \to 0$ regions are at the intersections of $x_l \to 1$
regions, so these collinear singularities also enhance the soft
singularity of the density. The normalization constant is
fixed so that $\int dN = 1$ and is ${\cal N}_x = 1/(6\sqrt 2
\arcsin(1/\sqrt 3))$. One can, of course, change $1/\sqrt{1-x}$ to
$1/(1-x)^A$ for any A smaller than 1.

It is a simple matter to actually choose points with the density $\rho$
in Eq.~(\ref{rhonet}). Readers interested in the implementation details
may consult the notes \cite{beowulfnotes} that accompany the program
code.

I can comment here about the singularities of the integrand, $f$. If,
for the moment, we set the measurement function to 1, then the nominal
behavior of $f$ is $f \propto 1/x_j^3$ for $x_j \to 0$ and
$f \propto 1/(1-x_j)$ for $x_j \to 1$. This nominal behavior is what
one gets after performing the virtual loop integral, $\int\! dl_1$,
inside the
$\{\vec p_1, \vec p_2\}$ integral \cite{sterman}. These singularities
make the integration logarithmically divergent. Then the measurement
function that is included in $f$ is required to vanish in the two jet
limit, so that the integration becomes convergent. Now, in fact, if we
look at the singularities in the $\{\vec p_1, \vec p_2\}$ integral
before performing the virtual loop integral, the singularities are
worse than this nominal behavior. For this reason, the suppression of
the two jet region coming from the measurement function must be
suitably strong in order to obtain good convergence. It remains for
future research to arrange the calculation in such a way that the
nominal behavior of the integrand $f$ as a function of $\{\vec p_1,
\vec p_2\}$ is obtained.

\section{Sampling for {\it 2 to 2 (s)} scattering}
\label{sec:2to2s}

In this section, we consider the sampling method associated with {\it 2
to 2 (s)} scattering. We need to choose points $\vec  l
\equiv \vec  l_1$ appropriate to the following case: 

\begin{quote}
A virtual parton with momentum $\vec  l$ merges with a
virtual parton with momentum $\vec  l_2 + \vec  l_3 -
\vec  l$ to produce a virtual parton with momentum $\vec l_2 +
\vec  l_3$. This virtual parton divides into two partons with 
momenta $\vec  l_2$ and $\vec  l_3$ that enter the final state. 
\end{quote}

\noindent
In this case, the scattering surface is the ellipsoid $|\vec  l| +
|\vec  l_2 + \vec  l_3 - \vec  l| = |\vec  l_2|+|\vec  l_3|$.
The density of points should be large on this surface. In order to
accomplish this, we use elliptical coordinates.

\subsection{Elliptical coordinates}

The elliptical coordinates $A_+,A_-,\phi$ are defined as follows.
First, define $\kappa$ by
\begin{equation}
2\kappa = |\vec  l_2 +\vec  l_3| .
\end{equation}
Now, define coordinates $A_\pm$ by
\begin{equation}
A_\pm = { 1 \over 2\kappa}\,
\left(
|\vec l| \pm |\vec l - \vec l_2 - \vec  l_3|
\right).
\end{equation}
Then
\begin{equation}
1 < A_+ 
,\hskip 1 cm 
-1 < A_- < 1.
\end{equation}
The coordinate $A_+$ is constant on elliptical surfaces, while $A_-$ is
constant on orthogonal hyperbolic surfaces. The scattering singularity is the ellipsoid
\begin{equation}
A_+ = S_+ ,
\end{equation}
where
\begin{equation}
S_\pm = {1 \over 2\kappa}\left(|\vec  l_2| \pm |\vec  l_3|\right).
\end{equation}

We need one more coordinate. Let $\phi$ be the azimuthal angle of $\vec
l$ in a coordinate system in which the $z$ axis lies in the direction of 
$\vec  l_2 + \vec  l_3$ and $\vec  l_2$ has zero $y$ component and a
positive $x$ component. To state this precisely, define unit vectors
\begin{eqnarray}
\vec n_z  &=&
{\vec  l_2 + \vec  l_3\over |\vec  l_2 + \vec  l_3|}
\nonumber\\
\vec n_y &=&
{ \vec  l_3 \times \vec  l_2 \over |\vec  l_3 \times \vec  l_2|}
\nonumber\\
\vec n_x &=& \vec n_y \times \vec n_z.
\label{ndef}
\end{eqnarray}
Then
\begin{equation}
\phi = \arctan\left(
\vec l\cdot \vec n_y
/\vec l\cdot \vec n_x
\right).
\end{equation}

The transformation from $A_+,A_-,\phi$ to $\vec  l$ is
\begin{equation}
\vec l =
{1\over 2} (\vec  l_2 + \vec  l_3)
+ l_T\,\cos\phi\ \vec n_x
+ l_T\,\sin\phi\ \vec n_y
+ z\, \vec n_z ,
\end{equation}
where
\begin{eqnarray}
l_T &=&\kappa
\left(A_+^2 - 1\right)^{1/2}
\left(1 - A_-^2\right)^{1/2}
\nonumber\\
z&=&\kappa\, A_+ A_- .
\end{eqnarray}

A straightforward calculation shows that the jacobian of the
transformation is given by
\begin{equation}
d\vec l = { dA_+ dA_- d\phi \over \rho_{AA\phi}} ,
\end{equation}
where
\begin{equation}
{ 1 \over \rho_{AA\phi}} = \kappa^3 (A_+^2 - A_-^2).
\label{rhoAAphi}
\end{equation}
The factor $(A_+^2 - A_-^2)$ is convenient. It provides weak
singularities in the density of points when $A_+ \pm  A_- \to 0$, which
corresponds to $\vec l \to 0$ and $\vec l - \vec  l_2 -\vec l_3 \to 0$.

If we sample points in the variables $\{A_+,A_-,\phi\}$ with a density 
\begin{equation}
\rho' = { dN \over dA_+\,dA_-\,d\phi},
\end{equation}
then the total density of points will be
\begin{equation}
\rho = \rho' \times \rho_{AA\phi},
\end{equation}
where $\rho_{AA\phi}$ is given in Eq.~(\ref{rhoAAphi}). We next address
the question of what to choose for $\rho'$.

\subsection{The choice of $A_+$, $A_-$ and $\phi$}

With what density
\begin{equation}
\rho' = { dN \over dA_+\,dA_-\,d\phi}
\end{equation}
should we sample points in the variables $\{A_+,A_-,\phi\}$? There is
no unique answer, but consider a density function of the form
\begin{equation}
\rho' = {\cal N}\,{ 1 \over 1-A_-^2 + \tau}\
{ 1 \over A_+(|A_+ - S_+|+\lambda)}.
\label{rho22s}
\end{equation}
Here $\tau$ and $\lambda$ are parameters to be specified and the
normalization ${\cal N}$ is given by
\begin{equation}
{ 1 \over {\cal N}} = 
{  2\pi \over \sqrt{1 + \tau}}\,
\ln\!\left(\sqrt{1 + \tau}+1 \over \sqrt{1 + \tau}-1 \right)\,
\left\{
{ 1 \over S_+ + \lambda}
\ln\!\left(S_+[S_+ - 1 + \lambda] \over \lambda\right)
+
{ 1 \over S_+ - \lambda}
\ln\left(S_+ \over \lambda\right)
\right\}.
\end{equation}

With this ansatz, there is a high density of points near the
scattering singularity surface, $A_+ = S_+$. The width of the peak is
$\lambda$. Thus $\lambda$ should be matched to the width of the
peak in the integrand. If the singularity surface is far from the line
$A_+ = 1$, the contour deformation is substantial and peak is broad. On
the other hand, if the singularity surface is near to the line $A_+
= 1$, the contour deformation is small and peak is narrow. I estimate
that the width of the peak in the integrand is of order $(S_+ -1)^2$
for $S_+$ near 1. For large $S_+$ it seems reasonable to choose a width
of order $S_+$. Thus I take
\begin{equation}
\lambda = { (S_+ -1)^2 \over S_+}.
\end{equation}

With the function $\rho'$ in Eq.~(\ref{rho22s}), there is an
enhancement of the density of points near $A_- = \pm 1$. 
This enhancement is built into the density in order to provide an extra
density of points near where the ellipsoid in Fig.~\ref{fig:singsd} is
tangent to the sphere. As explained in Sec.~\ref{sec:organization}, we
need a high density of points here when $\vec l_2$ and $\vec l_3$ are
nearly collinear, that is when $S_+$ is close to 1. I  arrange for this
by taking the parameter $\tau$ in Eq.~(\ref{rho22s}) to be
\begin{equation}
\tau = { S_+ -1 \over S_+}.
\end{equation}

It is easy to sample points $\{A_+,A_-,\phi\}$ with the density $\rho'$
in Eq.~(\ref{rho22s}) as explained in \cite{beowulfnotes}.

\section{Sampling for {\it 2 to 2 (t)} scattering}

In this section, we consider the sampling method associated with {\it 2
to 2 (t)} scattering. We need to choose points $\vec  l
\equiv \vec  l_1$ appropriate to the following case: 

\begin{quote}
A virtual parton with momentum $\vec  l_2 + \vec  l$ scatters from a
virtual parton with momentum $\vec  l_3 - \vec l$ by exchanging a
parton with momentum $\vec  l$. Partons with momentum $\vec  l_2$ and
$\vec  l_3$ emerge into the final state.
\end{quote}

In this case, there is a scattering singularity surface $|\vec  l_2
+ \vec  l| + |\vec  l_3 - \vec  l| = |\vec  l_2|+|\vec  l_3|$. There is
a singularity on this surface at the point where the momentum $\vec  l$
of the exchanged parton vanishes. The density of points should have a
matching singularity at this point.

\subsection{Elliptical coordinates}

As in the previous section, we use elliptical coordinates
$A_+,A_-,\phi$. The role of $\vec l$ in the previous section is now
played by $\vec l + \vec l_2$, so the formulas are a little different.
We again define $\kappa$ by
$
2\kappa = |\vec  l_2 +\vec  l_3|.
$
Now, define coordinates $A_\pm$ by
\begin{equation}
A_\pm = { 1 \over 2\kappa}\,
\left(
|\vec l_2 + \vec l| \pm |\vec  l_3 - \vec l |
\right).
\end{equation}
Then
$
1 < A_+ 
$ 
and 
$
-1 < A_- < 1.
$
Define
\begin{equation}
S_\pm = {1 \over 2\kappa}\left(|\vec  l_2| \pm |\vec  l_3|\right).
\end{equation}
The scattering singularity is the ellipsoid
\begin{equation}
A_+ = S_+
\end{equation}
while the soft exchange singularity is at 
\begin{equation}
\{A_+, A_-\}
=\{S_+, S_-\}.
\end{equation}
We need one more coordinate, an azimuthal angle $\phi$. We define unit
vectors $\{\vec n_x,\vec n_y,\vec n_z\}$ according to Eq.~(\ref{ndef}).
Then we define
\begin{equation}
\phi = \arctan\left(
(l_2 + \vec l\,)\cdot \vec n_y
/(l_2 + \vec l\,)\cdot \vec n_x
\right).
\end{equation}

The jacobian of the transformation from $\vec l$ to $\{A_+,A_-,\phi\}$
is again ${ 1 / \rho_{AA\phi}}$ where $\rho_{AA\phi}$ is given in
Eq.~(\ref{rhoAAphi}). If we sample points in the parameters
$\{A_+,A_-,\phi\}$ with a density $\rho'$ then the total density of
points will be
$
\rho = \rho' \times \rho_{AA\phi}.
$
We next address the question of what to choose for $\rho'$.

\subsection{The choice of $A_+$, $A_-$ and $\phi$}
\label{sec:choicet}

In this subsection we specify a choice for the density
\begin{equation}
\rho' = { dN \over dA_+\,dA_-\,d\phi}
\end{equation}
with which we sample in the variables $\{A_+,A_-,\phi\}$. We begin by
recognizing that we face two challenges. First, we would like to
have a concentration of points near the surface $A_+ = S_+$ with an
especially high density near $A_- = \pm 1$ if $S_+$ is near 1, as
discussed with respect to the sampling for {\it 2 to 2 (s)}. The second
challenge is to make $\rho'$ appropriately singular at $\{A_+,
A_-, \phi\} = \{S_+, S_-, 0\}$. We thus take $\rho'$ to have two parts
\begin{equation}
\rho' = \alpha_{2t}\, \rho_s + (1-\alpha_{2t})\rho_t,
\end{equation}
where $\alpha_{2t}$ is a fixed parameter with $0<\alpha_{2t} < 1$. We
take $\rho_s$ to be given by the density (\ref{rho22s}) for the sampling
for {\it 2 to 2 (s)}. Then we have met the first challenge. It remains
to design $\rho_t$ to address the second challenge.

The main idea is that there is a scattering singularity surface at $A_+
= S_+$, so that the integrand has a factor
\begin{equation}
{ 1 \over A_+ - S_+ + i \eta}
\end{equation}
where $\eta$ is the amount of deformation of the $S_+$ contour. The
amount of contour deformation vanishes at the soft exchange singularity
at $\{A_+, A_-, \phi\} = \{S_+, S_-, 0\}$ and, near to this point,
$\eta$ is proportional to the square of the distance to $\{S_+,
 S_-, 0\}$. Thus, on the surface $A_+ = S_+$ near to the soft
exchange  singularity, we can estimate
$\eta$ as $\omega^2$ where
\begin{equation}
\omega \equiv |A_- - S_-| + |\phi|/\pi .
\end{equation}
There is another factor of $1/\omega$ that arises from the propagator
of the exchanged parton after we take real-virtual cancellations into
account, as explained in detail in Ref.~\cite{beowulfPRD}. Thus the
absolute value of the integrand has a factor that can be estimated as
\begin{equation}
{ 1 \over \omega |A_+ - S_+|}
\label{desired}
\end{equation}
for $\omega \ll 1$ and $\omega^2 \ll |A_+ - S_+| \ll \omega$. We want
$\rho_t$ to have a singularity of the same nature, so that the
integrand divided by the density of points is singularity free. Thus we
take
\begin{equation}
\rho_t =
{ {\cal A} \over 
[|A_+ - S_+| + \beta_{2t} S_+ \omega^2]
[|A_+ - S_+| + \beta_{2t} \gamma_{2t}\, S_+ \omega] }.
\label{rhot}
\end{equation}
Here $\beta_{2t}$ and $\gamma_{2t}$ are fixed parameters and $\cal
A$ is a rather complicated function of $A_-$ and $\phi$ (described
below) that is of secondary importance. The main point is that $\rho_t$
behaves like (\ref{desired}), with cutoffs when $|A_+ - S_+|$ gets to
be smaller than $\omega^2$ or larger than $\omega$.

We have thus given $\rho_t$ an actual singularity at the point 
$\{A_+, A_-, \phi\} = \{S_+, S_-, 0\}$. This singularity in $\rho_t$ has
a special structure such that matches the structure of the integrand
$f$, so that $f/\rho_t$ is finite in the neighborhood of the soft
exchange singularity.  This point, which is important for the
convergence of the numerical integration, was described in some detail
in Ref.~\cite{beowulfPRD}. In that paper, however, the construction
was not as nice as one would like because the ellipsoidal coordinates
were not used and the ``ridgeline'' of $\rho$ was placed on the
tangent plane to the ellipsoid rather than on the ellipsoid itself. 

To sample points with the density $\rho_t$ of Eq.~(\ref{rhot}), we
sample first in $\phi$ with a density proportional to
$\ln^2(\pi\gamma_{2t}/|\phi|)$. Then, with $\phi$ chosen,  we sample in
$A_-$ with a density proportional to $\log(\gamma_{2t}/\omega)/\omega$.
Finally, having chosen $\phi$ and $A_-$, we sample in $A_+$ with a
density proportional to the right hand side of Eq.~(\ref{rhot}).
Taking into account the normalization conditions, this gives the
result in Eq.~(\ref{rhot}) with
\begin{eqnarray}
{ 1 \over {\cal A}}&=&
{ 4\pi 
\over S_+\beta}\
[\ln^2(\gamma_{2t}) + 2 \ln(\gamma_{2t}) + 2]\
{ 1 \over \gamma_{2t} -\omega}
\nonumber\\
&&\times
\left\{
1 + 
{ 1 \over 2\ln(\gamma_{2t}/\omega)}
\ln\!\left(
{ S_+ - 1 + \beta_{2t} S_+ \omega^2 \over 
  S_+ - 1 + \beta_{2t} \gamma_{2t}\, S_+ \omega}
\right)
\right\}
\nonumber\\
&&\times
\left\{
1 - 
{1\over 2\ln^2\!\left(\pi\gamma_{2t}/|\phi|\right)}
\left[\ln^2\!\left(
{\gamma_{2t} \over 
  1 + S_- + |\phi|/\pi}
\right)
+ 
\ln^2\!\left(
{\gamma_{2t} \over 
  1 - S_- + |\phi|/\pi}
\right)\right]
\right\}.
\end{eqnarray}
The function ${\cal A}$ may be a bit complicated, but it is quite
benign. In particular, ${\cal A}$ is not singular or zero anywhere
provided that $ \gamma_{2t} >  \omega$ holds everywhere. This requires
that $\gamma_{2t} > 3$.

\section{Sampling for {\it 2 to 3} scattering}

In this section, we consider the sampling method for the third type of
scattering singularity surface. We need to choose points $\vec l
\equiv \vec l_1$ appropriate to the following case: 

\begin{quote}
A virtual parton with momentum $\vec l$ collides with a
virtual parton with momentum $-\vec l$ to produce a final state
with partons carrying momenta $\vec l_2$, $\vec l_3$, and $-\vec
l_2 - \vec l_3$. 
\end{quote}

\noindent
In this case, the scattering surface is the sphere $2|\vec l|
 = |\vec l_2|+|\vec l_3| + |\vec l_2 + \vec l_3|$. The
density of points should be large on this surface. There is no special
point embedded in the surface where the density of points needs to be
singular. However, the density should be enhanced near the point of
this sphere where it intersects the ellipsoid in Fig.~\ref{fig:singsd} or
the narrower of the two ellipsoids in Fig.~\ref{fig:singse}.

Since the scattering singularity surface is a sphere, we use spherical
polar coordinates $\{r,\cos\theta,\phi\}$ to describe $\vec l$. We
choose the $\theta = 0$ axis along a certain direction $\vec P_C$. In
the case (d) illustrated in Fig.~\ref{fig:virtuald}, we define $\vec
P_C = - \vec p_3$. In the case (e) illustrated in
Fig.~\ref{fig:virtuale}, we define 
$\vec P_C = \vec p_1$ if $|\vec p_1| > |\vec p_3|$ or 
$\vec P_C = - \vec p_3$ if $|\vec p_3| > |\vec p_1|$.

The jacobian of the transformation from $\vec l$ to
$\{r,\cos\theta,\phi\}$ is
\begin{equation}
d\vec l = { dr\,d\cos\theta\,d\phi \over \rho_{r\theta\phi}},
\end{equation}
where
\begin{equation}
\rho_{r\theta\phi} = { 1 \over r^2}.
\end{equation}
If we sample points in the variables $\{r,\cos\theta,\phi\}$ with a
density
\begin{equation}
\rho'= { dN \over dr\,d\cos\theta\,d\phi} ,
\end{equation}
then the total density of points will be 
\begin{equation}
\rho = \rho' \times \rho_{r\theta\phi} .
\end{equation}

We now address the question of how to choose $\rho'$. Our analysis
follows closely that in Sec.~\ref{sec:choicet}. The main idea is that
there is a scattering singularity surface at $r = S$, where
\begin{equation}
S = {\textstyle {1 \over 2}}\,
(|\vec l_2|+|\vec l_3| + |\vec l_2 + \vec l_3|).
\end{equation}
Thus the integrand has a factor
\begin{equation}
{ 1 \over(r/S) - 1 + i \eta} ,
\end{equation}
where $\eta$ is the amount of deformation of the $r/S$ contour. The
amount of contour deformation vanishes quadratically as 
$\{r,\cos\theta\}$ approaches $\{S_C,1\}$, where
\begin{equation}
S_C = |\vec P_C|.
\end{equation}
The point $\{S_C,1\}$ is the point $\vec l = \vec P_C$. That is,
$\vec l = - \vec p_3$ in Fig.~\ref{fig:singsd} or in
Fig.~\ref{fig:singse} when $|\vec p_3| > |\vec p_1|$. Thus, on the
surface $r = S$, the amount of deformation $\eta$ can be estimated as
$\Delta^2$ where
\begin{equation}
\Delta = {\textstyle {1 \over 3}}\,
\sqrt{(1-S_C/S)^2 + (1 - \cos \theta)}.
\end{equation}
Thus the absolute value of the integrand has a factor that can be
estimated as
\begin{equation}
{ 1 \over |(r/S) - 1|}
\label{desiredagain}
\end{equation}
for $\Delta \ll 1$ and $\Delta^2 \ll |(r/S) - 1| \ll \Delta$. We want
$\rho'$ to have a singularity of the same nature, so that the
integrand divided by the density of points is singularity free.
Furthermore, we would like $\rho'$ to have an extra factor of
$1/\Delta^2$ to cancel the factor from the singularity associated with
the narrow ellipsoid in  Fig.~\ref{fig:virtuald} as we approach $\theta
= 0$. Thus we take
\begin{equation}
\rho' =
{  {\cal B} \over 
\Delta\,
\left(|(r/S) - 1| + \alpha_3\Delta^2 \right)
\left(|(r/S)  - 1| + \alpha_3\Delta \right)
},
\label{rho23}
\end{equation}
where $\alpha_3$ is a fixed parameter and
where ${\cal B}$ is a rather complicated but nonsingular function that
gets the normalization right:
\begin{equation}
{\cal B}
=
{\alpha_3 (1-\Delta) \over 18\pi S}
\left[\ln\left(
{ 1 \over \Delta^2}\,
{ 1 + \alpha_3\Delta^2 \over 1 + \alpha_3\Delta}
\right)
\right]^{-1}\
\left[\ln\left(
1 + {2 \over (1 - S_C/S)^2}
\right)
\right]^{-1}.
\end{equation}

\section{Sampling for {\it 2 to 1} scattering}

In this section, we consider the sampling method for the fourth type of
scattering singularity surface:

\begin{quote}
A virtual parton with momentum $\vec l$ combines with a
virtual parton with momentum $\vec l_2 - \vec l$ to produce an
on shell parton with momentum $\vec l_2$ that enters the final state.
\end{quote}

\noindent
As mentioned in Sec.~\ref{sec:organization}, the {\it 2 to 1} scattering
singularity surface is exceptional in that there is no singularity
except for the two singular points at $\vec l = 0$ and
$\vec l_2 - \vec l = 0$. Typically a {\it 2 to 1} scattering subdiagram
is part of a {\it 2 to 2 (s)} or {\it 2 to 2 (t)} scattering subdiagram
and the two singularities are provided for by the {\it 2 to 2 (s)} or
{\it 2 to 2 (t)} sampling methods. The exception is in the case of a
self-energy subgraph that is connected to a final state parton. In this
case, the {\it 2 to 2 (t)}, {\it 2 to 2 (s)}, and {\it 2 to 3} sampling
methods do not apply and we need an explicit {\it 2 to 1} sampling
method. Thus we apply a  {\it 2 to 1} sampling method only in the case
of a self-energy subgraph connected to a final state parton.

We will choose $\vec l$ with a density that is a linear combination
of  five densities:
\begin{equation}
\rho(\vec l) = 
\alpha_{1a}\, \rho_a(\vec l) 
+ \alpha_{1b}\, \rho_b(\vec l)
+ \alpha_{1a}\, \rho_a(\vec l - \vec l_2)
+ \alpha_{1b}\, \rho_b(\vec l  - \vec l_2)
+ (1-2\alpha_{1a} - 2 \alpha_{1b})\, \rho_c(\vec l).
\end{equation}
Here $\alpha_{1a}$ and $\alpha_{1b}$ are fixed positive parameters
with $(1-2\alpha_{1a} - 2 \alpha_{1b})$ also positive. 

For $\rho_a(\vec l)$ we take
\begin{equation}
\rho_a(\vec l) = { |\vec l_2| \over 4\pi}\,
{ 1 \over \vec l^{\,2}}
{ 1 \over \left(|\vec l|+ |\vec l_2|\right)^2}.
\label{rhoa}
\end{equation}
The density $\rho_b$ is a simple variation on this with $|\vec l_2|$
replaced by a scale $M$:
\begin{equation}
\rho_b(\vec l) = { M \over 4\pi}\,
{ 1 \over \vec l^{\,2}}
{ 1 \over \left(|\vec l|+ M\right)^2}.
\label{rhob}
\end{equation}
Here $M = (|\vec l_2| + |\vec l_3| + |\vec l_2 + \vec
l_3|)/3$, where the momenta of the particles
in the final state are $\vec l_2$, $\vec l_3$, and $-\vec l_2-\vec l_3$.
The $\rho_a$ and $\rho_b$ terms provide singularities at $\vec l^2 =
0$ and $(\vec l_2-\vec l)^2 = 0$.

The $\rho_c$ term provides a non-singular density of points in the
neighborhood of the collinear line $\vec l = x \vec l_2$ with $0 < x <
1$. For $\rho_c(\vec l)$ we take
\begin{equation}
\rho_c(\vec l) = 
{ \beta_1 \gamma_1^2 |\vec l_2| \over 2\pi}
{ 1 \over 
\sqrt{\beta_1^2 + (x - 1/2)^2}\,
\left(\sqrt{\beta_1^2 + (x - 1/2)^2}+ \beta_1\right)
\left(\vec l_T^{\,2} + \gamma_1^2 \vec l_2^{\,2} \right)^2
}.
\label{rhoc}
\end{equation}
Here $\beta_1$ and $\gamma_1$ are fixed parameters and $x$ and $\vec
l_T$ are defined by writing
\begin{equation}
\vec l = x \vec l_2 
+ \vec l_T,
\label{ellTdef}
\end{equation}
where $\vec l_T$ is the part of $\vec l$ that is orthogonal to
$\vec l_2$.

\section{Conclusions}

In a next-to-leading order calculation of three jet cross sections and
similar observables in electron-positron annihilation, a given graph
leads to several contributions to the cross section. Each of these
contributions has the form of a measurement function times a quantum
amplitude times a complex conjugate amplitude, as in
Fig.~\ref{fig:cutdiagrams}. We must integrate the sum of these
contributions over three loop momenta.  In this paper, I have
presented a method for sampling the space of loop momenta in order to
perform the integrations by the Monte Carlo method.

The main organizing principle for this sampling is to construct the
density of integration points as a sum and to have one or more terms in
this sum correspond to each possible three parton final state for the
graph.

A cut graph with a three parton final state has a virtual loop.  We
are thus led to consider a simple problem in quantum mechanics: where in
the space of the loop momentum are the singularities for two body to
$n$ body scatterings? The generic answer is that the singularities lie
on ellipsoidal surfaces.  This leads us to choose elliptical/hyperbolic
coordinates for the loop momentum space. For the most part, we avoid
letting the integrand be truly singular at these surfaces by giving the
loop momentum a suitable imaginary part according to the contour
deformation recipe of Ref.~\cite{beowulfPRD}.  Nevertheless, it is
helpful to make the density of integration points large near these
ellipsoidal surfaces.

We are left with special points on the scattering singularity surfaces
where the contour deformation vanishes, so that the integrand is
actually singular.  This singularity corresponds to soft parton
exchange and is significant in gauge field theories.  We have seen how
to give the density of points a singularity structure that matches
that of the integrand as one approaches a soft singularity.

The sampling method that has been described here is far from optimal.
It is, however, at least reasonably systematic, and it gives good
results for three jet observables that do not get significant
contributions from parton final states that are near to two jet
configurations.

\acknowledgements
This work was supported in part by the U.S.~Department of Energy.


\begin{thebibliography}{99}

\bibitem{ERT} 
R.~K.~Ellis, D.~A.~Ross and A.~E.~Terrano,
Nucl.\ Phys.\ B {\bf 178}, 421 (1981).

\bibitem{KN} 
Z.~Kunszt, P.~Nason, G.~ Marchesini and B.~R.~Webber 
in {\it Z Physics at LEP1}, Vol.~1, edited by B.~Altarelli,
R.~Kleiss ad C.~Verzegnassi (CERN, Geneva, 1989), p.~373

\bibitem{beowulfPRL} 
D.~E.~Soper,
Phys.\ Rev.\ Lett.\ {\bf 81}, 2638 (1998)
[hep-ph/9804454].

\bibitem{beowulfPRD}
D.~E.~Soper,
Phys.\ Rev.\ D {\bf 62}, 014009 (2000)
[hep-ph/9910292].

\bibitem{beowulfcode} 
D.~E.~Soper, {\it beowulf} Version 1.1,
http://zebu.uoregon.edu/$\sim$soper/beowulf/.

\bibitem{beowulfnotes}
D.~E.~Soper, {\it Beowulf 1.1 Technical Notes},
http://zebu.uoregon.edu/$\sim$soper/beowulf/.

\bibitem{KS} 
Z.~Kunszt and D.~E.~Soper,
Phys.\ Rev.\ D {\bf 46}, 192 (1992).

\bibitem{sterman} 
G.~Sterman,
Phys.\ Rev.\ D {\bf 17}, 2773 (1978).



\end{thebibliography}
\end{document}